\documentclass{article}

\PassOptionsToPackage{numbers, compress}{natbib}

\usepackage{microtype}
\usepackage{graphicx}
\usepackage{subfigure}
\usepackage{booktabs} 
\usepackage{bbm}
\usepackage{amsmath}
\usepackage{amssymb}

\usepackage{svg}

\usepackage{verbatim}
\usepackage{hyperref}

\usepackage[preprint]{neurips_2023}

\usepackage[utf8]{inputenc} 
\usepackage[T1]{fontenc}    
\usepackage{hyperref}       
\usepackage{url}            
\usepackage{booktabs}       
\usepackage{amsfonts}       
\usepackage{nicefrac}       
\usepackage{microtype}      
\usepackage{xcolor}         

\usepackage{multirow}
\usepackage{boldline} 
\usepackage{amsfonts, bm}
\usepackage{amsmath}
\usepackage{amssymb}

\usepackage{graphicx}
\usepackage{subfigure}

\usepackage{verbatim}
\usepackage{svg}
\usepackage{adjustbox}

\usepackage{amssymb}
\usepackage{pifont}
\usepackage{arydshln}

\author{%
  Soheil Hor\\
  Department of Electrical Engineering\\
    Stanford University\\
  Stanford, CA 94305 \\
  soheilh@stanford.edu \\
    \And
    Amin Arbabian\\
      Department of Electrical Engineering\\
    Stanford University\\
  Stanford, CA 94305 \\
  arbabian@stanford.edu \\
}

\title{Sense, Predict, Adapt, Repeat:\\
A Blueprint for Design of New Adaptive AI-Centric Sensing Systems}

\begin{document}

\maketitle

\begin{abstract} 

As Moore's Law loses momentum, improving size, performance, and efficiency of processors has become increasingly challenging, ending the era of predictable improvements in hardware performance. Meanwhile, the widespread incorporation of high-definition sensors in consumer devices and autonomous technologies has fueled a significant upsurge in sensory data. Current global trends reveal that the volume of generated data already exceeds human consumption capacity, making AI algorithms the primary consumers of data worldwide. 
To address this, a novel approach to designing AI-centric sensing systems is needed that can bridge the gap between the increasing capabilities of high-definition sensors and the limitations of AI processors.

This paper provides an overview of efficient sensing and perception methods in both AI and sensing domains, emphasizing the necessity of co-designing AI algorithms and sensing systems for \textit{dynamic} perception. The proposed approach involves a framework for designing and analyzing dynamic AI-in-the-loop sensing systems, suggesting a fundamentally new method for designing adaptive sensing systems through inference-time AI-to-sensor feedback and end-to-end efficiency and performance optimization.

\end{abstract}

\section{Introduction}
\label{section:intro}

In recent years, there have been significant advancements in sensor technology, making sensors more capable and accessible. Today's sensors can capture the world around us in unprecedented high detail, with even more significant improvements expected in the future. Examples include autonomous cars equipped with a suite of cameras, lidars and radars that can map and navigate in complex outdoor environments in real-time, and Augmented Reality headsets that rely on real-time fusion of data from an array of cameras and inertial sensors to map, locate and track humans and their surrounding environment.

\begin{figure}[ht]
    \centering
    \includegraphics[]{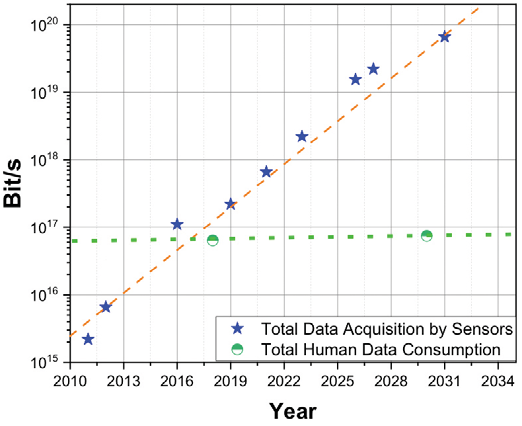}
    \caption{World sensor data trend and projection. Source:  \cite{srcdecadalplan}}

    \label{fig:bits-per_year}
\end{figure}

However, the high resolution of modern sensors comes at a cost: an enormous volume of raw sensory data. An autonomous car on average generates equivalent of 10 years of human generated data volume every day \cite{krzanich2016data}. This trend will only be only reinforced by introduction of new ``spatial computer''s and smart wearables to the consumer space. As shown in Figure \ref{fig:bits-per_year}, the amount of data generated by sensors has already surpassed total human data consumption capacity by an order of magnitude and is expected to grow exponentially in the years to come \cite{srcdecadalplan}. As a result, the majority of sensory data today and in the future will remain unseen by humans. 

Fortunately, the combination of high-speed communication links and powerful cloud computing has made it possible to train and run large-scale Deep Neural Networks and multi-modal foundational models that can convert high-dimensional sensory data into actionable information. 
Yet, sending the vast influx of raw sensory data to the cloud demands exponential advancements in computation power and communication infrastructure—challenges exacerbated by the slowdown of Moore's Law.
Using an optimistic figure for total energy efficiency of communication links {(1-10nJ/bit)}, and data-rates projected by \cite{srcdecadalplan} ($10^{20}$bits/s), the communication power costs of transmitting the global sensory data to the cloud can add up 100s to 1000s of Gigawatts of power. Using an estimated energy cost of around 0.1\$/kWh, this figure translates to over 2.4 Billion USD per day only in energy costs of communications. One solution to this problem is to reduce the cost of data communication to the cloud by performing ``at the edge'' processing and distillation of the raw data into abstracted information before transmission to the cloud. 

Local processing at the edge comes with its own unique set of challenges. Compared to the cloud, an edge device typically is more constrained in terms of both peak and average power consumption and its access to resources, such as memory and compute units. These limitations make on-device deployment of large-scale, high-performance machine learning models impractical, especially for real-time perception applications. 

Furthermore, attaining hardware capable of meeting performance requirements in specific applications is only part of the challenge; minimizing power consumption remains a fundamental hurdle. In the case of handheld devices or autonomous cars, the power and energy demands of sensing and computation units significantly impact device battery life. This renders it impractical to fully utilize both high-resolution sensors and high-performance machine learning algorithms. Simultaneously achieving optimum efficiency and performance in a resource-limited sensing system necessitates a paradigm shift in the design approach for both sensors and the data processing methods.


Conventional methods in designing resource-aware neural networks and sensing systems are often constrained, as they tend to overlook two fundamental design considerations:

First, the sensor and the perception models are typically modeled as black-boxes, each being designed and optimized Independently.
In resource-aware neural network design literature, the sensor and related signal conditioning blocks are often abstracted as an input stream of data with fixed and known properties. Therefore, the sensor-specific trade-offs and settings are usually simulated using simple data transforms and augmentations, which are not typically a good representation of the sensor's behavior and the real world environments. On the other hand, in system design literature, neural networks are typically modeled as a black-box consumer of sensory data with unpredictable responses to changes in sensing parameters and settings. One goal of our proposed approach is end-to-end optimization of the entire sensing system through co-design of neural networks and the sensing system. 

Secondly, sensing systems and their associated neural networks are commonly crafted in a `static' manner. This implies that once the perception neural networks are trained and optimized, they are presumed to remain fixed during the inference phase. This static approach yields a system designed for either average or peak performance and efficiency requirements, potentially falling short in scenarios with fluctuating (and dynamic) levels of difficulty.

As a conceptual example consider the dynamic complexity of the perception tasks performed by an autonomous car. Although the sensors and computational units of the vehicle should be designed to work in the most complicated scenarios (e.g., driving in a crowded city or in adverse and extreme weather conditions), most of the time the complexity of the perception task (and therefore the required computational power), is expected to be significantly lower, (e.g., being stuck in a traffic jam). Therefore the perception stack of an autonomous car should be designed to adapt to the dynamic complexity of the real-world driving environments.
Designing a system that can dynamically adjust its resource consumption based on the instance-level difficulty of the inference task is one of the main goals of the proposed approach.

This paper proposes a approach called ``end-to-end closed-loop active inference'' for efficient sensing on resource-constrained systems. Our emphasis on model-to-sensor feedback and dynamic closed-loop sensing offers a new approach to design of resource-constrained sensing systems. We provide an overview of static resource-aware approaches in section 2 and explore dynamic resource-aware approaches in sections 3 and 4. In section 5, we present our proposed approach, which aims to overcome limitations and provide a more effective solution for resource-aware design of adaptive AI-centric sensing systems.

\begin{figure*}[ht]
    \centering
    \includegraphics[width=\columnwidth]{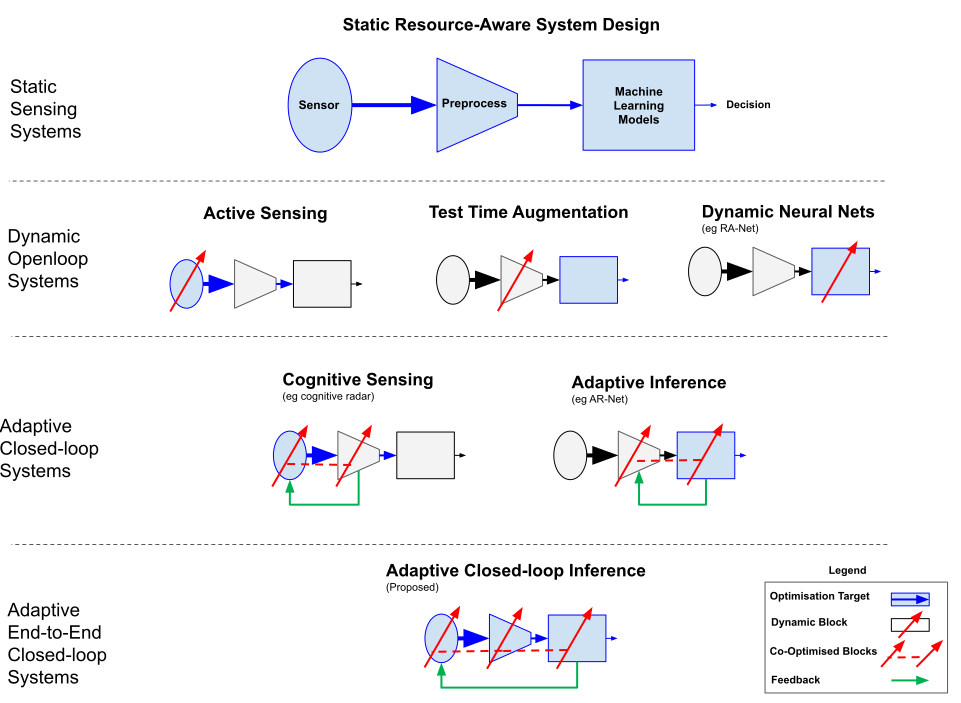}
    \caption{Resource-Aware Sensing Landscape}
    \label{fig:field_overview}
\end{figure*}

\section{Static Resource-Aware Sensing}
In this section we discuss the static approach to resource-aware sensing system design from two complementary perspectives: Sensor design and Neural Network design.

\subsection{Static Resource-Aware Sensor Design}
In recent years, high resolution sensors have become more powerful and accessible. However, the sensing system's ability to store, transfer and process sensory data remains a limiting factor in most edge processing applications. In this section, we introduce two main classical approaches to resource-aware sensor design that focus on making the sensor's raw data stream easier to manage.

\subsubsection{Compressive Sensing}
Compressive sensing is an umbrella term for all methods that rely on compressing the data directly at the source (sensor) to reduce the cost of data storage and transfer. The concept was first introduced by Donoho et al. \cite{donoho} in 2006. The simplest variant of compressive sensing methods relies on a custom design of the sensor, which can map the sensed environment into a set of maximally independent measurement vectors generated using a pseudo-random number generator or a pre-calculated dictionary.

Compressive sensing methods assume that the true information content recorded by the sensor is much smaller than the raw data volume. The promise of compressive sensing methods is to extract the information content of the sensory signal from a much smaller number of measurements than required by the Nyquist theorem \cite{donoho}. However, these methods usually rely on complex reconstruction (decompression) algorithms needed further down the data processing pipeline, are sensitive to the choice of reconstruction parameters, are typically designed in an static manner, and involve simplified linear assumptions that do not typically hold for perception tasks relying on complex Deep Neural Networks.

The simplest variant of compressive sensing methods can be defined with a fixed set of sampling bases, making it a static sensor design approach. However, modifying the sampling scheme to allow the compressive sensor to dynamically change the sampling basis between measurements can transform a compressive sensor into a dynamic, resource-efficient sensor. This dynamic approach is the basis of cognitive sensing literature discussed in section 3.

\subsubsection{In-Sensor Processing}
In-sensor processing has been proposed as an alternative to conventional off-sensor processing to reduce power consumption and memory usage \cite{inpixel}. Data filtering and transformation at the source can be much more efficient in terms of power and memory usage. A common approach in the literature is to embed preprocessing filters and transforms, and even first layers of neural networks, directly into the sensor itself (or individual pixels as  in \cite{scamp}), which can significantly reduce the amount of data that needs to be transmitted and processed elsewhere.

\cite{inpixel} provides an overview of advantages and challenges in in-sensor and new-sensor processing highlighting the potential of new 3D integration schemes in embedding memory and processing units into the sensor. The simplest examples of such approaches include pixels with nonlinear sensing functions. In particular, logarithmic pixel sensors have been shown to reduce power consumption and improve dynamic range compared to traditional linear pixel sensors \cite{murman_log}. 

\subsection{Static Resource-Aware Neural Network Design}
The remarkable success of neural networks in high-dimensional data analysis is largely due to the availability of large datasets and the advancement of cloud computing. However, the pursuit of ever-larger models to gain marginal improvements in accuracy has become impractical for resource-limited applications due to the high memory and computational requirements.

To tackle this issue, several methods have been proposed in the literature that aim to reduce the memory and computational requirements of neural networks, either by modifying the network's compute graph after the initial training phase or by resource-aware training of the neural networks. However, in both of these scenarios, neural network compression is assumed to be an ``offline'' optimization step, which results in a static computational graph during the inference phase. In this section, we introduce some of the most common methods used for (static) neural network compression.

\subsubsection{Pruning}
Pruning techniques are one of the most well-known methods for reducing the memory and computational requirements of a neural network. Neural network pruning is motivated by an empirical observation that in most applications, a large portion of weights in deep neural networks naturally converge to values close to zero \cite{han}.

In pruning approaches, the smaller weights in a pre-trained neural network are rounded down to zero, expecting minimum effect on the performance of the network. By designing custom neural network accelerators, it is possible to skip loading and computation of the zero weights, which results in a significant improvement in memory, computation, and power consumption at the inference time. It has been shown that by using custom losses during the training phase or iterative train-prune-train cycles, one can amplify the number of prune-ready weights while maintaining the original accuracy of the neural network. For a recent survey on the subject please see \cite{li_survey}.

While quite effective in reducing the memory requirements of neural networks, pruning methods often rely on structural changes to the neural networks. This typically results in a non-uniform computation graph which can be challenging to efficiently run on neural network accelerators and GPUs. Therefore, pruning methods alone might not be sufficient for static neural network efficiency optimisation \cite{osifeko2020artificial}. Quantization methods on the other hand are a family of methods specifically designed to make implementation of neural networks on neural network accelerators and GPUs more efficient as described in next section.

\subsubsection{Quantization}
Another empirical observation with neural networks is their resilience to small random variances in the inputs or even the network's weights. This property of neural networks can be exploited by reducing the number of bits used to describe each weight or activation from floating-point numbers (typically 32 or 64 bits) to integers (commonly 16 or 8 bits). Using integers to describe weights and activations in neural networks not only decreases the memory requirement of neural networks but also makes it possible to design custom efficient hardware for multiplication and addition operations that can significantly reduce the latency and power consumption of neural nets during the inference phase \cite{sze_survey}.

An extreme case of neural network quantization is in the case of single-bit quantization or binarization approaches, which reduce multiplication and addition operations to binary operations that can be efficiently done by simple digital gates instead of complex CPUs or GPUs (\cite{xnor_net}. 
While very effective in reducing both memory and computation requirements of neural networks, reducing the redundancies of neural networks through quantization can result in the network becoming less robust specially to adversarial attacks \cite{gorsline2021adversarial}. For a recent review on different quantization methods and their effect please see \cite{gholami2022survey}.

The effect of both quantization and pruning methods on the performance of neural networks is very task and network-dependent. Meaning that the initial choice of the network structure and its training can greatly affect the maximum efficiency and performance achievable by quantization and pruning methods. Resource-aware Neural Architecture Search is one of the approaches that complements the effect of the mentioned methods by finding the optimum network architecture during the training phase.

\subsubsection{Neural Architecture Search}
While pruning and quantization methods can significantly reduce the resource requirements of neural networks, they are still limited by the performance and efficiency limits of their original pre-trained neural network. Resource-aware Neural Architecture Search (NAS) takes network compression one step further by searching for the optimum network structure that can provide the best performance with minimum resource consumption \cite{white2023neural}.

Neural networks designed by NAS techniques have outperformed hand-designed neural networks in terms of both efficiency and performance across a wide range of applications. For example, the Efficient-Net family has outperformed large hand-designed neural networks in terms of accuracy per FLOPS for image classification tasks \cite{effnet}, and hardware-aware-NAS has outperformed small hand-designed models in terms of memory use, latency, and model size on micro-controllers and other edge devices (for a survey on the topic please see  \cite{benmeziane2021comprehensive}.

In theory, NAS methods can achieve the optimum neural network structure limited only by the ability of the architecture search algorithms to cover the space of possible neural net structures. However, even an ideal NAS method is limited by the fundamental assumption that the network compute graph is the same for all instances during the inference phase. Dynamic approaches explained in the next section can overcome this limitation by changing the network structure dynamically during the inference phase.

\section{Dynamic Resource-Aware Sensing}

\subsection{Dynamic Resource-Aware Sensor design: Cognitive Sensing}

Cognitive sensing is a dynamic system design paradigm that aims to tailor a sensor's parameters to the inferred context of the environment \cite{osifeko2020artificial}. By selecting the optimal sensing modality and parameters, the sensory signal's information content is maximized. Although cognitive sensing broadly refers to methods that emulate the human sensing-perception-action cycle in sensing systems, it has mainly been used in two specific contexts: cognitive radars, which control the sensor to minimize interference and effect of clutter (\cite{haykin2006cognitive}), and cognitive IoT networks(\cite{wu2014cognitive}, \cite{matin2020towards}) which extract actionable insights from distributed sensory data. Our proposed approach, although aligns with the broad definition of cognitive sensing, takes a fundamentally different approach to the sensing problem in the sense that it aims to simultaneously optimize a black-box AI model's efficiency and performance through AI-to-sensor feedback. This approach represents a conceptual shift in cognitive system design from an \textit{AI-assisted human-centric sensing} approach to an \textit{AI-centric closed-loop sensing} approach.

\subsection{Dynamic Resource-Aware Neural Network Design}

\subsubsection{Test-Time Augmentation}
Test Time Augmentation (TTA) (\cite{tta_cnn}, \cite{understanding_tta}) refers to a collection of techniques used to enhance the performance of a static neural network by transforming or augmenting the sensory signal during the inference phase. Unlike conventional data augmentation methods which are typically applied during the training phase, and aim to increase the size of training sets and improve neural network robustness to variations in input, TTA is used during inference.

The fundamental concept of TTA is to apply various transformations to the neural network's input during inference, then aggregate the prediction outcomes to boost the model's accuracy. TTA can increase the information about the correct class and minimize mis-classifications caused by overfitting on a specific representation of the input signal. TTA can be used to motivate the use of sensor adaptation techniques to improve the performance of deep neural networks. However, TTA has limited applications in resource-aware sensing system design because it assumes a fixed neural network architecture and fixed data format and volume for the raw sensory inputs.

\subsubsection{Adaptive Inference}
Adaptive inference is an emerging field in machine learning that focuses on the adaptation of neural networks to dynamic environments during inference \cite{graves2016adaptive} \cite{passalis2020efficient}. This is driven by the realization that neural networks are typically designed assuming a fixed and unbiased test set, whereas in practice the model may encounter samples with varying levels of inference difficulty. Adaptive inference relies on the promise that by adjusting the neural network's complexity during inference, one can reduce resource consumption for easier instances while requiring more resources for more difficult cases. By dynamically tuning the model's complexity with the inference task's difficulty, it is possible to reduce resource consumption with minimal impact on the model's performance. In practice, adaptive inference is typically performed by adaptive early exiting methods \cite{laskaridis2021adaptive}, \cite{RANet},\cite{passalis2020efficient}. However, as it will be discussed in the next section, the concept can be extended to other degrees of freedom of a dynamic neural network (for a survey on dynamic neural networks, see \cite{dynamicNN_survey}).

Adaptive inference methods offer a framework for designing dynamic closed-loop methods that prioritize the performance of a deep neural network. However, these methods do not inherently take into account the limitations and trade-offs of the sensor and other components of the system. As a result, the adaptive inference methods are typically solely focus on optimisation of the number of parameters and FLOPS of the perception networks.

Our proposed approach seeks to optimize both the inference performance, and the system's end-to-end efficiency, through AI-to-sensor feedback. By doing so, we aim to achieve a more comprehensive and efficient sensing system that takes into account the entire system's performance.

\section{Adaptive End-to-End Closed-Loop Inference}
In the preceding sections, various methods were explored, each offering valuable perspectives on resource-aware design in sensing systems. However, none of the mentioned approaches are suitable for design of end-to-end closed-loop inference systems.

\begin{figure}[ht]
    \centering
    \includegraphics[width=\columnwidth]{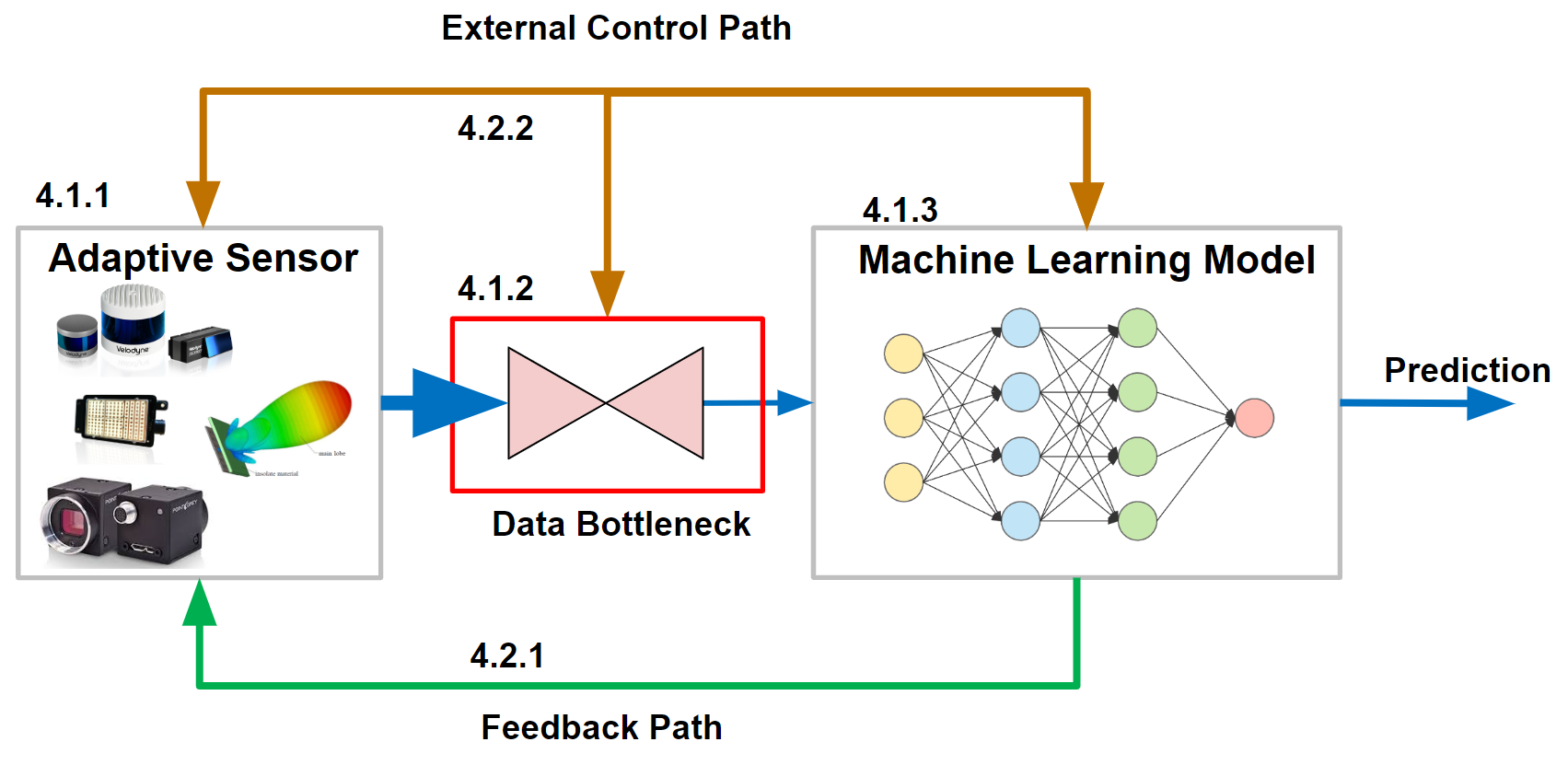}
    \caption{Adaptive System Overview}
    \label{fig:overview}
\end{figure}

Figure~\ref{fig:overview} presents a conceptual block diagram of a closed-loop system and its different components. To maintain generality, we represent the components of a sensing system as a sensor, a (Deep) Perception Model, and a conceptual ``data bottleneck'' block positioned in between. The main data path (shown in blue) and two control paths (shown in orange and green) describe the interactions between these primary blocks within the conceptual sensing system. In this section, we provide insights into various considerations and design approaches regarding data and control paths.

Throughout this discussion, we draw relevant parallels between the high-level concepts introduced here and a more technical paper on the subject, exemplifying the application of the proposed adaptive closed-loop inference approach in the context of radar waveform adaptation for pedestrian gait analysis \cite{waveform_adapt}.

\subsection{Data-path}
The conceptual data path depicted in Figure~\ref{fig:overview} comprises three fundamental blocks that collectively characterize a sensing system at the highest level: The (adaptive) sensor, the data-bottleneck block, and the (tunable) Perception Model.

\subsubsection{Adaptive Sensor}
In this paper, we adopt a comprehensive definition of the term ``sensor'' which accommodates a wide range of applications and resource-limited systems. Within our context, an adaptive sensor refers to any sensor equipped with one or more controllable knobs (referred to as ``controllable Degrees of Freedom''), allowing dynamic adjustments to modify the extent or precision of the sensor's readings along various axes of information. For instance, cameras with advanced lens systems can fine-tune parameters like resolution and field of view, while RGB-D cameras can seamlessly enable or disable depth-sensing and color imaging modalities. This inclusive definition covers both passive and active sensors. However, active sensors, such as mm-wave radars and scanning lidars, hold particular significance due to their extra controllable degrees of freedom and the unique challenges they encounter in actively probing and interacting with their sensing environment.

Controllable Degrees of Freedom in adaptive sensors manifest across various domains, including time (managing frame rate and total framing time across different time scales), space (controlling spatial resolution and field of view in 2D or 3D space), spectrum (managing bandwidth and center frequency in active sensing systems like radars), and power levels (regulating the transmit signal power in active sensors like lidars or radars).

At the highest level, one can conceptualize multi-modal, multi-view, and multi-sensor systems as a ``meta-sensor''. In such scenarios, the meta-sensor's Degrees of Freedom enable it to select or prioritize specific modalities, views, or sensors, based on the system's state. This approach allows us to abstract the entire sensing stack of complex sensing systems (e.g., autonomous cars) into an adaptive sensor with a defined set of states, each optimized for a specific operation mode. For instance, an autonomous car might have high-speed mode, prioritizing long-range sensors and high frame rates, and low-speed mode, focusing on spatial resolution for precise localization in closer spaces.

As an example, the adaptive sensor described in \cite{waveform_adapt} is as an FMCW radar system designed for human gait analysis applications, with the unique capability of measuring micro-motions of the human body. The sensor can dynamically adjust the velocity (Doppler) measurement resolution of the radar by modifying the number of transmitted radar pulses (chirps) on the fly. However, achieving enhanced velocity resolution involves an increase in raw data volume, creating a trade-off between performance and efficiency.

In summary, our definition of adaptive sensors encompasses a wide range of controllable Degrees of Freedom, enabling dynamic adjustments in various domains, ultimately leading to more versatile and resource-efficient sensing systems across a multitude of applications.

\subsubsection{Data Bottleneck}
One of the key justifications for dynamic perception is the presence of inevitable data bottlenecks in every sensing system. As explained in Section \ref{section:intro}, processing the data generated by high-resolution sensors is often impractical, especially on resource-limited edge devices. Consequently, data must undergo selection, transformation, or compression before being processed by the perception model. To simplify our model for the data path, we abstract all factors that could limit a system's data rate into a single bottleneck block.

These limiting factors encompass various considerations, such as memory, space, time/latency, power, and data rate, either enforced by the system's physics (e.g., maximum data rate of I/O ports) or driven by practical design choices (e.g., limiting the maximum laser beam power of a lidar system for eye safety). Additionally, the limitations of the perception model itself, including maximum model size, latency, and power consumption of the neural net accelerator, can also be incorporated into the bottleneck block.

These limitations introduce trade-offs between different degrees of freedom within the data path, initially established by the sensor. For example, the frame resolution and frame rate of a camera are typically considered to be two independent design parameters. However, by imposing constraints on the maximum data rate or minimum signal-to-noise ratio per pixel, the average frame rate and accessible frame resolution of the camera become interdependent.

The bottleneck block, by defining the limits and constraints on the Degrees of Freedom of the adaptive sensor, serves as the foundation for defining the state space of the data path. However, a complete definition of the data path and its state space requires understanding of the perception model itself and its optimized design for each state.

\subsubsection{Adaptive Perception Models}
The proposed closed-loop inference approach differentiates itself from classical closed-loop sensing systems by assuming that a Deep Neural Network (employed as the perception model) is the primary, and often sole, consumer of the sensory data. As a result, the perception model not only plays a crucial role in defining the optimization objective of the sensing system, but also becomes one of the most critical blocks to be optimized

However, typical static perception models lack three essential properties: they do not support different input types (e.g., modalities) and formats, they cannot handle dynamic changes in input availability (e.g., dealing with missing input modalities), and they lack the ability to dynamically tune the network's computational complexity.

Recent advancements in multi-modal neural network design hold substantial promise in addressing the first two aforementioned limitations. For instance, Bachmann et al. have demonstrated the effectiveness of modality-specific encoders paired with task-specific decoders for multi-modal and multi-task image analysis \cite{bachmann2022multimae}. Similarly, Girdhar et al. have presented the possibility of creating scalable general-purpose multi-modal embedding spaces. These spaces allow the formation of modality-agnostic data representations suitable for various recognition and perception tasks \cite{girdhar2023imagebind}. These approaches exemplify ways in which perception models can efficiently consume data from different modalities with dynamic availability. Nonetheless, devising encoders and decoders readily adaptable for perception-at-the-edge applications remains an ongoing challenge.

When addressing particular dynamic degrees of freedom (e.g., spatial extent and resolution, and time extent and resolution), specific strategies can be employed to convert a static neural network into an input-adaptive version with minimal computational overhead or complexity. An illustrative approach involves replacing conventional pooling blocks within convolutional neural networks with adaptive pooling blocks \cite{adaptive_pool}, and substituting fully connected layers with equivalent 1x1 convolutional layers.

By employing this approach, an off-the-shelf static neural network with a fixed input size and resolution can be converted into a fully-convolutional neural network, virtually agnostic to input size and resolution variations. Networks with depth-wise separable convolutions as their building blocks (e.g., Mobile-Net \cite{mobilenet} and Shuffle-Net\cite{shufflenet}) are particularly suited for this purpose due to their small field of view in each convolution stack, enabling processing of inputs with arbitrary size with minimal changes to the network. Notably, certain networks inherently possess resilience to changes in the input's resolution and extent \textit{in time}. Illustrative examples include the Temporal Segment Network \cite{TSN} and its variants like ActionNet \cite{actionNet}

An alternative approach to employing these methods involves utilizing Neural Architecture Search (NAS) to train an ensemble of static networks, each corresponding to a specific state of the data-path. While this approach demands substantial training and inference resources, it ensures the optimal operating point (in terms of both performance and efficiency) for each input state.

Furthermore, an intriguing avenue in perception model design involves incorporating adaptive routing or other dynamic network tuning methods. These techniques introduce an additional degree of freedom to balance performance and efficiency within the network which can broaden the range of operational states of the data path. For further insights, readers are encouraged to explore a comprehensive survey on this subject by Han et al. \cite{dynamicNN_survey}.

A fundamental premise of the proposed closed-loop approach is that the data path, particularly the perception model, has undergone prior optimization for maximum efficiency and performance across its various states (utilizing the static optimization techniques mentioned in the preceding sections). As a result, the exclusive task of the closed-loop system is to optimize the transitions between these data-path states. Our proposed methodology achieves this through the integration of a control path, which encompasses both AI-to-sensor feedback and external dynamic control signals. This aspect is elaborated upon in the subsequent section.

\subsection{Control path}

\subsubsection{Feedback Signal}
It can be argued that the feedback path is the most important piece of the proposed approach to close-loop sensing system design. A distinctive feature that sets this approach apart from other closed-loop methods discussed earlier is its explicit goal for the feedback path to oversee the complete state of the data-path, encompassing the sensor, the black-box (tunable) perception model, and the associated bottleneck blocks. This choice gives rise to a distinct set of challenges and considerations.

An essential factor in crafting the feedback method design is the selection of the optimization objective. Typically, this objective entails formulating a reward function (or loss function) based on the desired performance metrics (such as accuracy, precision, or various error rate goals), diverse cost parameters (such as memory, power consumption, latency, bandwidth), and their respective significance.

For instance, \cite{waveform_adapt} employs a linear combination of classification accuracy (performance) and the relative data volume across the complete data-path as its reward function. This approach parallels the common practice in adaptive inference studies (e.g., \cite{ar_net}), where linear loss functions are defined in relation to accuracy and the average FLOPS of the adaptive model.

The optimization method used to determine the optimal adaptation strategy varies based on the complexity of the optimization function. While Reinforcement Learning, as shown in \cite{waveform_adapt}, offers a general approach adaptable to any reward function, simpler reward functions can be optimized using conventional convex optimization methods or even intuitive rule-based policies.

As an example, if the importance of a particular term within the reward function greatly outweights the others, it is often possible to frame the adaptation goal as the well-recognized min-max problem. In perception systems that value performance (accuracy) more than efficiency, this results in a straightforward approach: finding the state with the least resource consumption while maximizing performance (accuracy). This intuitive concept can guide the development of simple, yet effective rule-based adaptation methods or even solving the adaptation problem using supervised learning approaches.

The feedback control path serves as the last essential component in crafting an end-to-end closed-loop sensing system capable of adapting to the dynamics of its sensory input. Yet, within real-world sensing systems, the perception stack is often developed in coordination with other subsystems. The subsequent section delves into the external control path, which conceptually represents the interface between the proposed closed-loop sensing system, and external control signals (originating from other units within a larger system or even human operators).

\subsubsection{External Control Signal}
\label{section:system_constraints_path}

As previously outlined, interaction between the perception sub-system and other sub-systems (such as path planning, power control, and user interface) within a larger entity (like an autonomous car) is often necessary. The external control path, as delineated in this section, encapsulates the interplay among the data-path, the feedback-path, and external control units that exist beyond the closed-loop perception system.

Within our proposed model, external control over the data-path and its constituents (including the adaptive sensor, perception model, and data bottleneck) can be executed by constraining or broadening the adaptation state-space. For instance, a human user might opt to confine the system to states that ensure a specific threshold of perception robustness and accuracy. Alternatively, the power management unit of an autonomous system could dynamically establish a cap on the maximum power consumption permitted for the perception subsystem.

Likewise, a linkage between the feedback path and external control signals can be established by manipulating the reward function of the feedback path. This encompasses the ability to manually or automatically regulate the significance of performance and efficiency within the reward function, depending on the overall resources accessible to the perception subsystem.

The following section delves further into the challenges and factors to consider in relation to the suggested closed-loop inference approach.

\section{Challenges and Considerations}

Similar to any novel paradigm, the proposed approach introduces its unique set of challenges and considerations. Within this section, we offer a high-level description of the crucial factors involved in designing adaptive end-to-end closed-loop systems, along with insightful perspectives on potential solutions for addressing each challenge.

\paragraph{The Need for New Dynamic Datasets and Simulation Environments} 
The core motivation behind embracing adaptive perception as an alternative to classical static approaches is to enable the system to adapt to the ever-changing dynamics of the environment. Thus, it becomes imperative to capture a comprehensive and realistic representation of the diverse dynamic factors influencing the perception task. A significant challenge lies in simultaneously obtaining generalizable representations of both the environmental dynamics, and the dynamic sensing systems, while considering their interactions through feedback mechanism. To address this challenge, potential solutions include employing post-processing techniques (e.g., downsampling) to simulate sensor adaptation, integrating data from multiple static sensors to emulate a dynamic sensor, or utilizing simulators capable of modeling both the environment and the adaptive sensing system's ``digital twin''s.

\paragraph{The Need for New Benchmarks and Figures of Merit} 
In the evaluation of adaptive sensing systems, two primary considerations stand out. 

One crucial aspect, often overlooked even in Active Inference literature, is the isolation of the impact of the \textit{test-set dynamics} (hereby referred to as system's ``adaptation potential''), from the influence of the adaptation policy of a closed-loop sensing system (referred to as ``adaptation regret''). To quantify the adaptation potential, a viable solution involves first evaluating the performance of open-loop and closed-loop adaptive Oracles (optimal algorithms with complete knowledge of both the environment's dynamics and the system) on the test-set. By comparing adaptive strategies to adaptive Oracles, it becomes possible to disentangle the effects of adaptation potential and regret.

Another challenge lies in defining new figures of merit for adaptive systems. The current approach in Adaptive Inference typically relies on reporting a single quantitative measure, such as best efficiency/FLOPS for a fixed performance or best accuracy for a fixed resource budget. However, the true value of adaptation emerges from significantly expanding the operation space of perception systems in the real-world, enabling a more favorable overall performance-efficiency trade-off that may not be achieved at the point of maximum performance nor maximum efficiency. To address this, a potential solution is to define a set of objective and subjective measures that compare the resource-vs-performance curves of different systems across their entire operation space.

\paragraph{Challenges in Training Policy Networks} 
Determining the ground truth for defining the optimum adaptation policy is a challenging task. It requires a comprehensive understanding of potential short-term and long-term outcomes for each action and sensing system state. 
{Consequently, training policy networks in many scenarios can only be accomplished through Reinforcement Learning approaches, which are renowned for their difficulty in training. A potential solution is to integrate data-driven and model-based approaches or use heuristics from traditional closed-loop system design to formulate the closed-loop perception problem within the context of supervised or unsupervised learning.}

\paragraph{Challenges in Closed-loop Inference} 
Inference of closed-loop perception systems introduces its own unique set of challenges and considerations. This includes positive feedback effects, where the feedback policy reinforces the errors of the perception model and vice-versa, resulting in a loop of incorrect states and predictions. Additionally, stability issues can arise due to a noisy feedback path, causing the system to switch between different states without any specific efficiency or performance gain. On the other hand, sub-optimal performance may occur due to overly conservative (i.e., slow) feedback responses. As these challenges are also prevalent in other classical closed-loop systems, a potential solution lies in the intersection of local linear approximations of neural networks (e.g., LIME \cite{lime}), and classical methods in robust control, which offer guarantees on the robustness and stability of closed-loop systems.

\paragraph{Inherent Costs of Adaptation}
Designing a dynamic system comes with inherent costs compared to its static counterpart. For instance, changing the state of the data-path may necessitate interrupting the data stream, reloading control buffers of dynamic sensors, loading new weights for the perception neural network, or re-routing parts of the data stream. This extra complexity imposes inherent costs in terms of latency (e.g., for sensor reconfigurations and stream interruptions), power consumption (e.g., energy cost of memory access for weight updates), memory (e.g., storing weights for several states into memory), space (e.g., requiring extra routing and control hardware for the control path), and compute (e.g., the additional computational load of the policy network).

Depending on the application, these costs can be significant, especially since the adaptation task itself, in some cases, might require a network of similar complexity as the primary perception task. These challenges necessitate a fundamentally new approach to designing sensing systems, one that allows for efficient dynamic operation of the system.

Examples of such approaches include parameter sharing between the perception model and the policy network, generating weights on-the-fly for each state, and the design of sensors and processors capable of efficiently handling dynamic data loads under different settings.

\section{Conclusion}

In summary, this white paper examines static, dynamic, and adaptive closed-loop approaches to AI-centric sensing system design, offering a conceptual abstraction for end-to-end co-design of AI algorithms and sensing systems. We introduced a novel design framework emphasizing optimization through AI-to-sensor feedback, followed by an exploration of challenges and opportunities in real-world implementation. As the demand for AI-at-the-edge sensing systems continues to rise, we aspire for this white paper to serve as a catalyst, inspiring further research into design and implementation of adaptive AI-centric sensing systems

\bibliography{main}

\begin{thebibliography}{10}

\bibitem{krzanich2016data}
Brian Krzanich.
\newblock Data is the new oil in the future of automated driving.
\newblock {\em Intel Editorial}, 2016.

\bibitem{srcdecadalplan}
SRC.
\newblock Decadal plan for semiconductors, January 2021.

\bibitem{donoho}
David~L Donoho.
\newblock Compressed sensing.
\newblock {\em IEEE Transactions on information theory}, 52(4):1289--1306,
  2006.

\bibitem{inpixel}
Gourav Datta, Souvik Kundu, Zihan Yin, Ravi~Teja Lakkireddy, Joe Mathai, Ajey~P
  Jacob, Peter~A Beerel, and Akhilesh~R Jaiswal.
\newblock A processing-in-pixel-in-memory paradigm for resource-constrained
  tinyml applications.
\newblock {\em Scientific Reports}, 12(1):14396, 2022.

\bibitem{scamp}
Laurie Bose, Piotr Dudek, Jianing Chen, Stephen~J Carey, and Walterio~W
  Mayol-Cuevas.
\newblock Fully embedding fast convolutional networks on pixel processor
  arrays.
\newblock In {\em Computer Vision--ECCV 2020: 16th European Conference,
  Glasgow, UK, August 23--28, 2020, Proceedings, Part XXIX 16}, pages 488--503.
  Springer, 2020.

\bibitem{murman_log}
Daisuke Miyashita, Edward~H Lee, and Boris Murmann.
\newblock Convolutional neural networks using logarithmic data representation.
\newblock {\em arXiv preprint arXiv:1603.01025}, 2016.

\bibitem{han}
Song Han, Huizi Mao, and William~J Dally.
\newblock Deep compression: Compressing deep neural networks with pruning,
  trained quantization and huffman coding.
\newblock {\em arXiv preprint arXiv:1510.00149}, 2015.

\bibitem{li_survey}
Zhuo Li, Hengyi Li, and Lin Meng.
\newblock Model compression for deep neural networks: A survey.
\newblock {\em Computers}, 12(3), 2023.

\bibitem{osifeko2020artificial}
Martins~O Osifeko, Gerhard~P Hancke, and Adnan~M Abu-Mahfouz.
\newblock Artificial intelligence techniques for cognitive sensing in future
  iot: State-of-the-art, potentials, and challenges.
\newblock {\em Journal of Sensor and Actuator Networks}, 9(2):21, 2020.

\bibitem{sze_survey}
Vivienne Sze, Yu-Hsin Chen, Tien-Ju Yang, and Joel~S. Emer.
\newblock Efficient processing of deep neural networks: A tutorial and survey.
\newblock {\em Proceedings of the IEEE}, 105(12):2295--2329, 2017.

\bibitem{xnor_net}
Mohammad Rastegari, Vicente Ordonez, Joseph Redmon, and Ali Farhadi.
\newblock Xnor-net: Imagenet classification using binary convolutional neural
  networks.
\newblock In {\em European conference on computer vision}, pages 525--542.
  Springer, 2016.

\bibitem{gorsline2021adversarial}
Micah Gorsline, James Smith, and Cory Merkel.
\newblock On the adversarial robustness of quantized neural networks.
\newblock In {\em Proceedings of the 2021 on Great Lakes Symposium on VLSI},
  pages 189--194, 2021.

\bibitem{gholami2022survey}
Amir Gholami, Sehoon Kim, Zhen Dong, Zhewei Yao, Michael~W Mahoney, and Kurt
  Keutzer.
\newblock A survey of quantization methods for efficient neural network
  inference.
\newblock In {\em Low-Power Computer Vision}, pages 291--326. Chapman and
  Hall/CRC, 2022.

\bibitem{white2023neural}
Colin White, Mahmoud Safari, Rhea Sukthanker, Binxin Ru, Thomas Elsken, Arber
  Zela, Debadeepta Dey, and Frank Hutter.
\newblock Neural architecture search: Insights from 1000 papers.
\newblock {\em arXiv preprint arXiv:2301.08727}, 2023.

\bibitem{effnet}
Mingxing Tan and Quoc Le.
\newblock Efficientnet: Rethinking model scaling for convolutional neural
  networks.
\newblock In {\em International conference on machine learning}, pages
  6105--6114. PMLR, 2019.

\bibitem{benmeziane2021comprehensive}
Hadjer Benmeziane, Kaoutar~El Maghraoui, Hamza Ouarnoughi, Smail Niar, Martin
  Wistuba, and Naigang Wang.
\newblock A comprehensive survey on hardware-aware neural architecture search.
\newblock {\em arXiv preprint arXiv:2101.09336}, 2021.

\bibitem{haykin2006cognitive}
Simon Haykin.
\newblock Cognitive radar: a way of the future.
\newblock {\em IEEE signal processing magazine}, 23(1):30--40, 2006.

\bibitem{wu2014cognitive}
Qihui Wu, Guoru Ding, Yuhua Xu, Shuo Feng, Zhiyong Du, Jinlong Wang, and Keping
  Long.
\newblock Cognitive internet of things: a new paradigm beyond connection.
\newblock {\em IEEE Internet of Things journal}, 1(2):129--143, 2014.

\bibitem{matin2020towards}
Mohammad~Abdul Matin.
\newblock {\em Towards Cognitive IoT Networks}.
\newblock Springer, 2020.

\bibitem{tta_cnn}
Guotai Wang, Wenqi Li, S{\'e}bastien Ourselin, and Tom Vercauteren.
\newblock Automatic brain tumor segmentation using convolutional neural
  networks with test-time augmentation.
\newblock In Alessandro Crimi, Spyridon Bakas, Hugo Kuijf, Farahani Keyvan,
  Mauricio Reyes, and Theo van Walsum, editors, {\em Brainlesion: Glioma,
  Multiple Sclerosis, Stroke and Traumatic Brain Injuries}, pages 61--72, Cham,
  2019. Springer International Publishing.

\bibitem{understanding_tta}
Masanari Kimura.
\newblock Understanding test-time augmentation.
\newblock In {\em International Conference on Neural Information Processing},
  pages 558--569. Springer, 2021.

\bibitem{graves2016adaptive}
Alex Graves.
\newblock Adaptive computation time for recurrent neural networks.
\newblock {\em arXiv preprint arXiv:1603.08983}, 2016.

\bibitem{passalis2020efficient}
Nikolaos Passalis, Jenni Raitoharju, Anastasios Tefas, and Moncef Gabbouj.
\newblock Efficient adaptive inference for deep convolutional neural networks
  using hierarchical early exits.
\newblock {\em Pattern Recognition}, 105:107346, 2020.

\bibitem{laskaridis2021adaptive}
Stefanos Laskaridis, Alexandros Kouris, and Nicholas~D Lane.
\newblock Adaptive inference through early-exit networks: Design, challenges
  and directions.
\newblock In {\em Proceedings of the 5th International Workshop on Embedded and
  Mobile Deep Learning}, pages 1--6, 2021.

\bibitem{RANet}
Le~Yang, Yizeng Han, Xi~Chen, Shiji Song, Jifeng Dai, and Gao Huang.
\newblock Resolution adaptive networks for efficient inference, 2020.

\bibitem{dynamicNN_survey}
Y.~Han, G.~Huang, S.~Song, L.~Yang, H.~Wang, and Y.~Wang.
\newblock Dynamic neural networks: A survey.
\newblock {\em IEEE Transactions on Pattern Analysis and Machine Intelligence},
  44(11):7436--7456, nov 2022.

\bibitem{waveform_adapt}
Soheil Hor, Mert Pilanci, and Amin Arbabian.
\newblock A data-driven waveform adaptation method for mm-wave gait
  classification at the edge.
\newblock {\em IEEE Signal Processing Letters}, 29:26--30, 2022.

\bibitem{bachmann2022multimae}
Roman Bachmann, David Mizrahi, Andrei Atanov, and Amir Zamir.
\newblock Multimae: Multi-modal multi-task masked autoencoders.
\newblock In {\em European Conference on Computer Vision}, pages 348--367.
  Springer, 2022.

\bibitem{girdhar2023imagebind}
Rohit Girdhar, Alaaeldin El-Nouby, Zhuang Liu, Mannat Singh, Kalyan~Vasudev
  Alwala, Armand Joulin, and Ishan Misra.
\newblock Imagebind: One embedding space to bind them all.
\newblock In {\em Proceedings of the IEEE/CVF Conference on Computer Vision and
  Pattern Recognition}, pages 15180--15190, 2023.

\bibitem{adaptive_pool}
Shu Liu, Lu~Qi, Haifang Qin, Jianping Shi, and Jiaya Jia.
\newblock Path aggregation network for instance segmentation.
\newblock In {\em Proceedings of the IEEE conference on computer vision and
  pattern recognition}, pages 8759--8768, 2018.

\bibitem{mobilenet}
Mark Sandler, Andrew Howard, Menglong Zhu, Andrey Zhmoginov, and Liang-Chieh
  Chen.
\newblock Mobilenetv2: Inverted residuals and linear bottlenecks.
\newblock In {\em Proceedings of the IEEE conference on computer vision and
  pattern recognition}, pages 4510--4520, 2018.

\bibitem{shufflenet}
Xiangyu Zhang, Xinyu Zhou, Mengxiao Lin, and Jian Sun.
\newblock Shufflenet: An extremely efficient convolutional neural network for
  mobile devices.
\newblock In {\em Proceedings of the IEEE conference on computer vision and
  pattern recognition}, pages 6848--6856, 2018.

\bibitem{TSN}
Limin Wang, Yuanjun Xiong, Zhe Wang, Yu~Qiao, Dahua Lin, Xiaoou Tang, and Luc
  Van~Gool.
\newblock Temporal segment networks: Towards good practices for deep action
  recognition.
\newblock In {\em European conference on computer vision}, pages 20--36.
  Springer, 2016.

\bibitem{actionNet}
Zhengwei Wang, Qi~She, and Aljosa Smolic.
\newblock Action-net: Multipath excitation for action recognition.
\newblock In {\em Proceedings of the IEEE/CVF conference on computer vision and
  pattern recognition}, pages 13214--13223, 2021.

\bibitem{ar_net}
Yue Meng, Chung-Ching Lin, Rameswar Panda, Prasanna Sattigeri, Leonid
  Karlinsky, Aude Oliva, Kate Saenko, and Rogerio Feris.
\newblock Ar-net: Adaptive frame resolution for efficient action recognition.
\newblock In {\em Computer Vision--ECCV 2020: 16th European Conference,
  Glasgow, UK, August 23--28, 2020, Proceedings, Part VII 16}, pages 86--104.
  Springer, 2020.

\bibitem{lime}
Marco~Tulio Ribeiro, Sameer Singh, and Carlos Guestrin.
\newblock " why should i trust you?" explaining the predictions of any
  classifier.
\newblock In {\em Proceedings of the 22nd ACM SIGKDD international conference
  on knowledge discovery and data mining}, pages 1135--1144, 2016.

\end{thebibliography}
\bibliographystyle{unsrt}

\end{document}